\documentclass[fleqn,10pt]{SelfArx} 

\usepackage{lipsum} 

\definecolor{color1}{RGB}{0,0,90} 
\definecolor{color2}{RGB}{0,20,20} 
\definecolor{color3}{RGB}{0,10,150} 

\usepackage{hyperref} 
\hypersetup{hidelinks,colorlinks,breaklinks=true,urlcolor=color3,citecolor=color1,linkcolor=color3,bookmarksopen=false,pdftitle={Title},pdfauthor={Author}}

\JournalInfo{Accepted by Geomagnetism and Aeronomy, 2019, V.59, No 7} 
\Archive{} 

\PaperTitle{
CYCLES ON THE SOLAR-TYPE STARS AND COOLER DWARFS
} 

\Authors{N.\,I.~Bondar\textsuperscript{1*}, M.\,M.~Katsova\textsuperscript{2**}, \fbox{M.\,A.~Livshits}\textsuperscript{3}
} 

\affiliation{\textsuperscript{1}\textit{
Crimean Astrophysical Observatory, Nauchny, Crimea, Russia
}} 
\affiliation{\textsuperscript{2}\textit{
Sternberg State Astronomical Institute, Lomonosov Moscow State University, Moscow, Russia
}} 
\affiliation{\textsuperscript{3}\textit{
Pushkov Institute of Terrestrial Magnetism, Ionosphere, and Radio Wave Propagation RAS, Troitsk, Moscow, Russia
}} 

\affiliation{\textsuperscript{*}e-mail: otbn@mail.ru}
\affiliation{\textsuperscript{**}e-mail: maria@sai.msu.ru}
\affiliation{Received: February 24, 2019}

\newcommand{\keywordname}{Keywords} 

\Abstract{
Features of the development of activity cycles in the solar-type stars and 
fast-rotating cool dwarfs have been considered for 65 stars observed in some decades. 
Cycles with duration of 7--18 years compared to the solar cycle were found for about 50\%\ 
of the studied stars. In cooler dwarfs with rotation periods of less than 5 days, 
cyclic changes in brightness occur on longer scales, up to 80 years. Activity of the highest 
level is produced on K dwarfs; their main cycles are long and have the highest amplitudes. 
Both old and young solar-type stars show a similar tendency in increasing the cycle length 
with a slower rotation. No evidence for a relation between the rotation period and duration 
of cycles was found for cool dwarfs with $P_\textsl{rot} < 5$\,days.
}

\begin{document}

\flushbottom 

\maketitle 

\tableofcontents 

\thispagestyle{empty} 

\section{INTRODUCTION}

The magnetic activity typical for the solar-type stars (spectral classes F8-K2, $T_\textsl{eff} = 6200–4500\;$K) and cooler K--M dwarfs with effective temperatures of up to $2200\;$K determines the diversity of variability observed in different wavelength ranges. The time scale of variability is in the range from seconds (fast flares) to tens of years (activity cycles similar to the solar magnetic cycle); there are also secular cycles of hundreds and thousands of years. On the Sun, manifestations of activity in different layers of the atmosphere are regulated by an 11-year cycle. The development of magnetic activity cycles is determined by the rotation of a star and its internal structure.

According to Nizamov et al.~(2017), cycles in G-type dwarfs are established from the epoch when their rotation period reaches 1.1 days, for K and M dwarfs these periods are longer -- 3.3 and 7.2 days, respectively. The stars of the corresponding spectral classes, rotating faster, demonstrate the saturated activity regime. 

A certain number of stars with cycles longer than the solar one are known, but most of the studied objects were investigated from the modern databases on a limited time interval; therefore, cycles of no longer than 10--15 years were found for these stars (Savanov, 2012; Suárez Mascareño et al., 2016; Boro Saikia et al., 2018). The results concerning stellar cycles require further specification and confirmation from the long-term systematic observations.

Oláh et al.~(2016) derived the series of data on time scales up to 36 years for 29 G-K stars from the Mount Wilson HK-Project and determined the duration of their dominant cycles (with a maximum amplitude) and short cycles.

The long-term variations in brightness of several dozens of cool K--M dwarfs were studied from photographic archives and broadband photometry (Bondar’, 1995, 2013; Alekseev and Bondar’, 1998; Alekseev and Kozhevnikova, 2017); for some of them the activity of photospheric starspots similar to the observed on the Sun has been found or suspected.

The task of our work is to form a sample of stars, whose cycles have been found from the data over several decades; to substantially extent the available photometric series for some G--M dwarfs by the results of recent observations; to refine or determine cycles of their photospheric activity. 

Besides, we will compare characteristics of dominant cycles of solar-type stars and cooler dwarfs obtained over long-term series, consider how they are related to rotation periods, reveal features of this connection for different type stars, that is important for the dynamo theory.

\section{G--M DWARFS WITH CYCLIC ACTIVITY\\ ON LONG-TERM INTERVALS}

The first results on activity cycles of F--M stars on or near the main sequence were obtained by Wilson (1978) within his monitoring program of 91 stars in the Ca II H and K emission lines. Based on observations in 1966--1977, for some stars there have been suspected cycles (trends) which require certain evidence from the longer data series. The program was continued in the framework of the HK-Project for 111 stars. In the paper by Baliunas et al.~(1995), which became the basis for studying magnetic activity of the solar-type stars, the cycles were searched for and classified from the data series which include the 12-year observations by O.\ Wilson and new data, and in whole cover the 25-year interval.

Oláh et al.~(2016) presented new data about chromospherical activity cycles for 29 HK-Project stars from the 36-year observation series. Most of these stars exhibit cycles with a duration of 7--18 years, i.e. which correspond to the sunspot cycle. Some stars, besides the main cycle, demonstrate one or several shorter cycles (multiple cycles), as well as long-term trends of 25 years or longer.

\begin{figure*}[!th] 
\centering
\includegraphics[width=\linewidth]{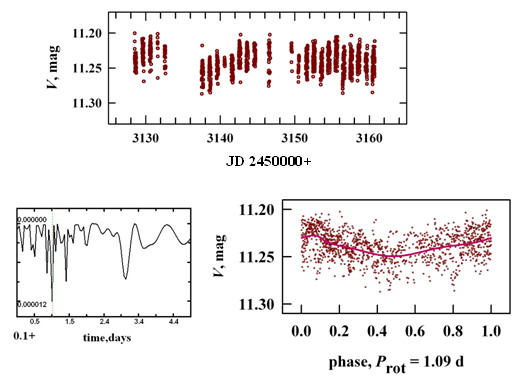}
\caption{
Search for the rotation period of V647 Her. Top panel: V -- light curve on the interval of JD2453128--2453160 (2004) from the SuperWASP data. Bottom panel: periodogram of the presented data, the peak corresponds to a period of 1.087 days (left), the phase fit of V-magnitudes with $P_\textsl{rot} = 1.087$\;days (right).
}
\label{Figure1}
\end{figure*}

For our purpose, we take into account the results of a search for cycles on cool dwarfs from the series longer than 20 years, i.e. an investigating time spans must be comparable with those for the solar-type stars in Oláh et al.~(2016).

We have used the results derived for 40 dK--dM stars from photographic data series (Bondar’, 1995, 2013), from photometric researches by Alekseev and Kozhevnikova (2017) and Messina et al.~(2002). For 9 stars (Bondar’ 2019) the available data series were added by data from photometric catalogs. Data from the Hipparcos catalog (\url{http://cds.u-strasbg.fr/cgi-bin/Dic-Simbad?HIP}) cover\\ the span of 1990–1993, data from the ASAS catalog (Pojmanski, 1997) were obtained in 2002–2009, data from SuperWASP (\url{https://wasp.cerit-sc.cz/form}) cover the interval between 2004–2008, data obtained in 2008\-–\-2018 were taken from the KWS database (Kamogata / Kiso / Kyoto Wide-field Survey, \url{http://kws.cetus-net.org/~maehara/VSdata.py?object}).

\begin{table*}[!th] 
\centering
\vspace*{0.5cm}
\caption{
G- and M-type dwarfs with active cycles determined from new photometric data series.
}
\vspace*{0.5cm}

\begin{tabular}{l|c|l|c|r|c|c|l|l}
Star & B-V & Sp type & $P_\textsl{rot}$, d & $P_\textsl{cyc}$, yr & $A_\textsl{cyc}$, mag & Time span & Variable type & Sources\\
\hline
BE Cet & 0.66 & G2.5 & 7.76 & 6.8 $\pm$ 0.1 & $<$ 0.04 & 1986–2018 & BY Dra & 2, 4, 5\\
DX Leo & 0.77 & G9 & 5.42 & 3.3 $\pm$ 0.1 & 0.03 & 1989–2018 & BY Dra & 1, 2, 4, 5\\
V1005 Ori & 1.37 & M0 & 4.40 & 20.4 $\pm$ 0.2 & 0.12 & 1976–2018 & BY Dra & 1, 2, 4, 7\\
DT Vir & 1.48 & M0 & 2.88 & 31.6 $\pm$ 0.4 & 0.15 & 1911–2018 & fl & 1, 2, 4, 6\\
V647 Her & 1.64 & M3.5 & 1.09 & 76.1 $\pm$ 0.6 & 0.30 & 1939–2018 & fl & 1, 2, 3, 4, 6\\
V577 Mon & 1.69 & M4.5 & 8.00 & 40 $\pm$ 0.4 & 0.50 & 1909–2018 & fl & 1, 2, 4, 6\\
DX Cnc & 2.06 & M6.5 & 0.46 & 2.7 $\pm$ 0.1 & 0.10 & 1984–2018 & fl & 2, 4, 6\\
V374 Peg & 1.26 & M3.5 & 0.44 & var & 0.10 & 1989–2017 & fl & 1, 2, 3, 4\\
HU Del & 1.65 & M4.5 & 0.35 & var & 0.32 & 1940–2017 & fl & 2, 4, 6\\
\end{tabular}
\vspace*{0.5cm}

{
Remarks. Spectral type, B-V and type of variability are taken from the SIMBAD astronomical database; numbers in column “Sources” mark databases Hipparcos (1), ASAS (2), SuperWASP (3), KWS (4); Messina and Guinan, 2002 (5), Bondar’, 2013 (6), Alekseev and Kozhevnikova, 2017 (7).
The type of stellar variability is marked according to the SIMBAD astronomical database: BY Dra-type stars (BY Dra), flare stars (fl).
}
\label{Table1}
\end{table*}

The list of stars and used data sources are given in Table~1. Two stars from the list -- BE Cet and DX Leo -- belong to G-dwarfs. On the interval between 1986--2000 Messina and Guinan (2002) found only short-term cycles for these stars. We obtained data series covering a time span twice as long, until 2018; this allows us to consider the stability of short cycles and to search for possible long-term cycles. The cooler stars in our sample have spectral classes M0V--M6.5V; for some of them the long-term photographic measurements of brightness were used from Bondar’ (1995). Two stars, V374 Peg and HU Del, are of interest as fast-rotating fully convective dwarfs. Light curves of V374 Peg from the photometry in 1998--2013 were studied by Vida et al.~(2016), but the authors found no cycles on this star. The parameters of cycles (the cycle length and amplitude) given in Table 1 were obtained as a result of the analysis of composite light curves formed by us, including recent data of 2017 and 2018. We made a search for the cycles from periodograms derived by the Yurkevich, Hartley and Scargle methods using the AVE program (\url{http://www.gea.cesca.es}) as described in (Bondar’ 2019).

Note that the errors in the $P_\textsl{cyc}$ values show the accuracy of their determination, but not in the duration of cycles, which may vary up to several years at different epochs.

Variations in light curves of V374 Peg and HU Del are not cyclic, and the parameter $A_\textsl{cyc}$ shows the magnitude of the maximum changes in their yearly mean brightness over the studied time interval. No long-term cycles were found on the BE Cet and DX Leo stars. We confirm that only the short cycles detected by Messina and Guinan (2002) have been found from the 30-year photometric series. Considering parameters of the cycle, we took information about the duration of cycles for V833 Tau (78 yr) on the interval of 1899–2009 from Bondar’ (2015) and for YZ CMi (27.5 yr) from data on the time span of 1926–2009 (Bondar’, 2019).

The rotation period of the star V647 Her was determined according to the SuperWASP data-base obtained with high temporal resolution. A search for the period was performed using the AVE program by the Hartley and Scargle methods on the JD2453128 – 2453160 (2004) interval containing 1292 points for periods of 0.1–5 days and 0.1–10 days. The periodogram in Fig.\ 1 (the right panel) shows the presence of a peak corresponding to a period of 1.09 days determined by the Scargle method (Scargle, 1982).

We also confirmed the rotation periods of the stars DT Vir and V374 Peg. According to our analysis of data from the ASAS catalog since 2003 to 2008, the rotation period of the star DT Vir $P_\textsl{rot} = 2.88\;$d is consistent with 
$P_\textsl{rot} = 2.85\;$d derived by Donati et al.~(2008); for V374 Peg, according to data in 2004 from the SuperWASP catalog, 
we found $P_\textsl{rot} = 0.448\;$d to agree with $P_\textsl{rot} = 0.446\;$d acquired by Vida et al.~(2016). 

\section{ROTATION PERIODS AND\\ PARAMETERS OF ACTIVITY CYCLES}

Here we consider 65 G–M dwarfs, for which a search for cycles has been performed on long-term intervals. A sample includes 36 stars studied from the photometric series and 29 stars, including the Sun, from the Mount Wilson HK-Project (Oláh et al., 2016). The number distribution of objects by spectral types is as follows: 14 G, 28 K, and 23 M dwarfs.

\begin{figure*}[!th] 
\centering
\vspace*{-0.9cm}
\includegraphics[width=\linewidth]{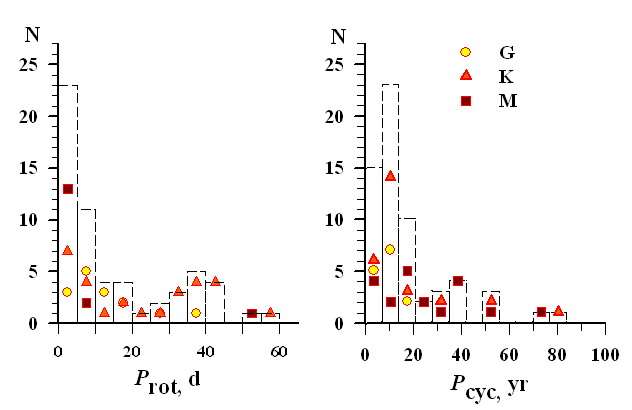}
\vspace*{0.2cm}
\caption{
Distribution of rotation periods and duration of cycles for stars from our sample.
}
\label{Figure2}
\end{figure*}

The rotational periods are known for 59 stars of the sample; cycles were detected or suspected for 62 stars. Fig.\ 2 (left panel) shows the distribution of periods in bins, each of them is equal to 5\ d, the spectral classes are marked with different symbols, the dashed line indicates the total number of stars in the bin. Most of G-type dwarfs (10 objects) have periods from 5 to 20 days and only the star HD 81809 (G2V) rotates slower than the Sun, with a period of 39.3 days. Fast-rotating dwarfs with periods of less than 5 days include 25 stars from the sample; mainly, these are cool dwarfs. Three M-stars have rotation periods of less than 0.5 day. 13 K--M dwarfs fall into the range of rotation periods of 30–60 days, the slowest ones are two objects HD 95735 (M2V) and HD 16160 (K3V) with rotation periods of 54 and 57 days, respectively.

The distribution on the length of cycles is shown in Fig.\ 2 (right panel), bins correspond to 7 years. The cycle durations of 50\%\ of stars are from 7 to 21 years that is comparable to the solar spot cycle. The long cycles, up to 80 years, are detected in fast-rotating K--M dwarfs. The length of dominant cycles for 15 stars is short, less than 7 years. The long-term brightness variations of fully convective M-dwarfs HU Del and V374~Peg with $P_\textsl{rot} <0.5$~d showed no certain periodicity. The light curves for some stars and the results of a search for the cycles are given in (Bondar', 2019).

\begin{figure*}[!th] 
\centering
\vspace*{-0.9cm}
\includegraphics[width=\linewidth]{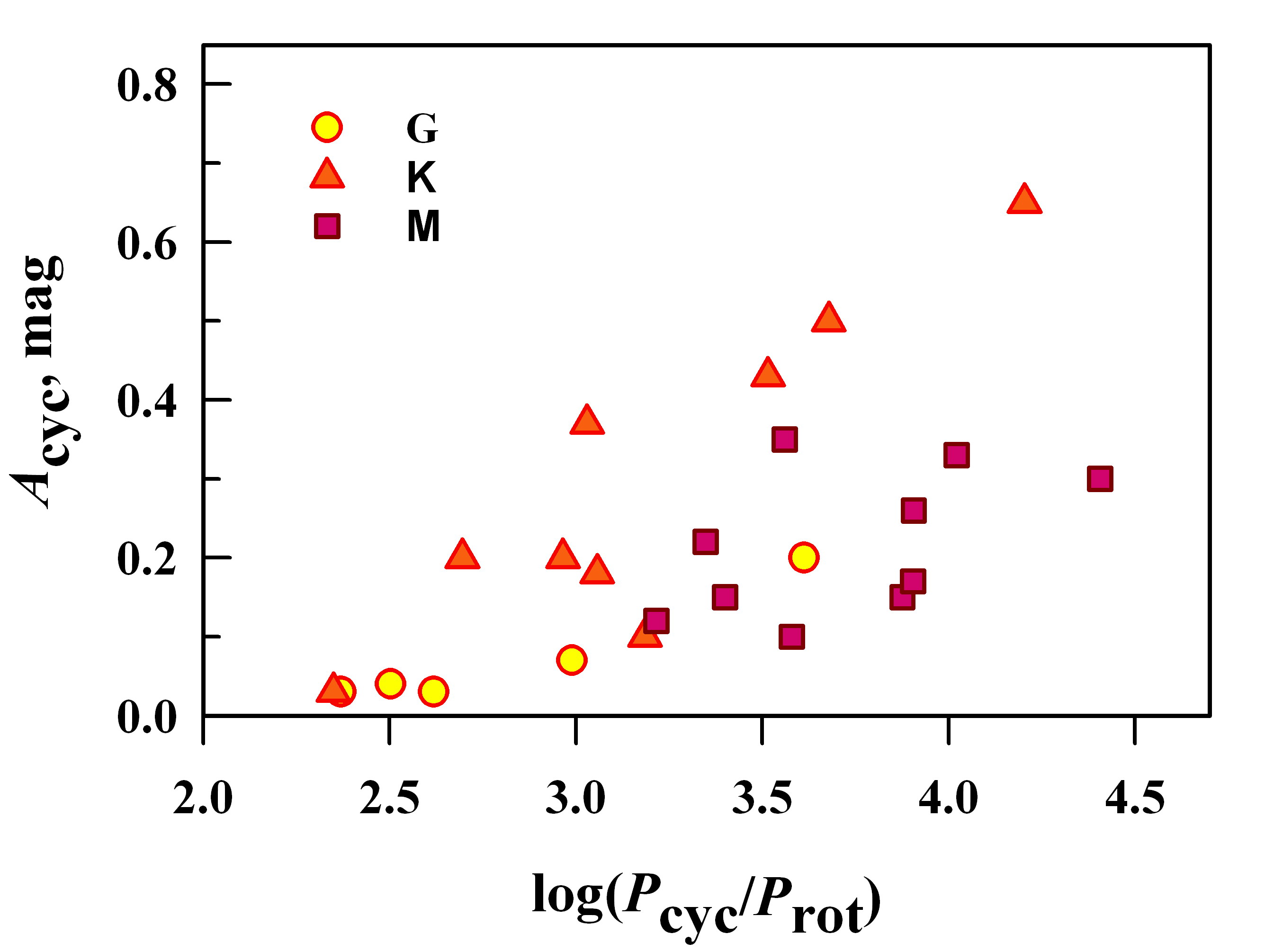}
\vspace*{0.2cm}
\caption{
The amplitude and duration of cycles of G–M dwarfs with different rotation periods.
}
\label{Figure3}
\end{figure*}

The amplitude of the dominant cycle in the solar-type stars does not exceed 0.2$^m$, and for the fast-rotating cool dwarfs it can increase up to 0.5$^m$ or more; that indicates on the large areas of active regions. As follows from Fig.~3, K-dwarfs with any rotation periods are distinguished from G and M dwarfs by such parameters as cycle duration and amplitude; this leads to the conclusion that their level of activity is the highest among dwarfs on or near the main sequence.

\section{FEATURES OF THE DEVELOPMENT\\ OF CYCLES IN G–M DWARFS}

The development of activity cycles in the solar-type stars mainly depends on their rotational velocity. On the diagram $P_\textsl{cyc}$ and $P_\textsl{rot}$ (Bondar’, 2019) stars are divided into two sequences near a value of $P_\textsl{rot} = 5\;$d. For stars with $P_\textsl{rot} > 5$~d, the cycle length increases if a rotation period becomes slower. This sequence is not homogeneous, since it includes objects with periods from several days to tens of days, i.e. having different ages. For the older stars, rotating slower than the Sun, the above tendency is less pronounced than for the young G and K dwarfs with $5 < P_\textsl{rot} < 20$~days. For these stars cycles of more than 22 years have not been detected yet, but among young K dwarfs there are stars with cycles of about 40 years. The Sun is located between groups of old and young stars.

The youngest stars with $P_\textsl{rot} <5$~d show no any definite connection between the rotation and cycle length. Baliunas et al.~(1996) derived a relation between the observed parameters $P_\textsl{cyc}$ and $P_\textsl{rot}$ and the theoretical dynamo number $D$ expressed as $P_\textsl{cyc}/P_\textsl{rot} \sim D^i$. Later, such a relation was considered in several papers for different samples of stars (Oláh et al., 2009; Savanov, 2012; Vida et al., 2014; Katsova et al., 2015; Suárez Mascareño et al., 2016). Vida et al.~(2014) have found for all the long and short cycles the value of the slope $i \sim 0.77$, which agrees with Baliunas et al.~(1996), but M dwarfs have been excluded from the analysis.

Suárez Mascareño et al.~(2016) and Savanov (2012) have considered only M stars; the stellar cycles were determined from observations on intervals of less than 10 years. The obtained large values of $i$ indicate that for the cooler dwarfs such a relation is weaker or absent.

\begin{figure*}[!th] 
\centering
\vspace*{-0.9cm}
\includegraphics[width=\linewidth]{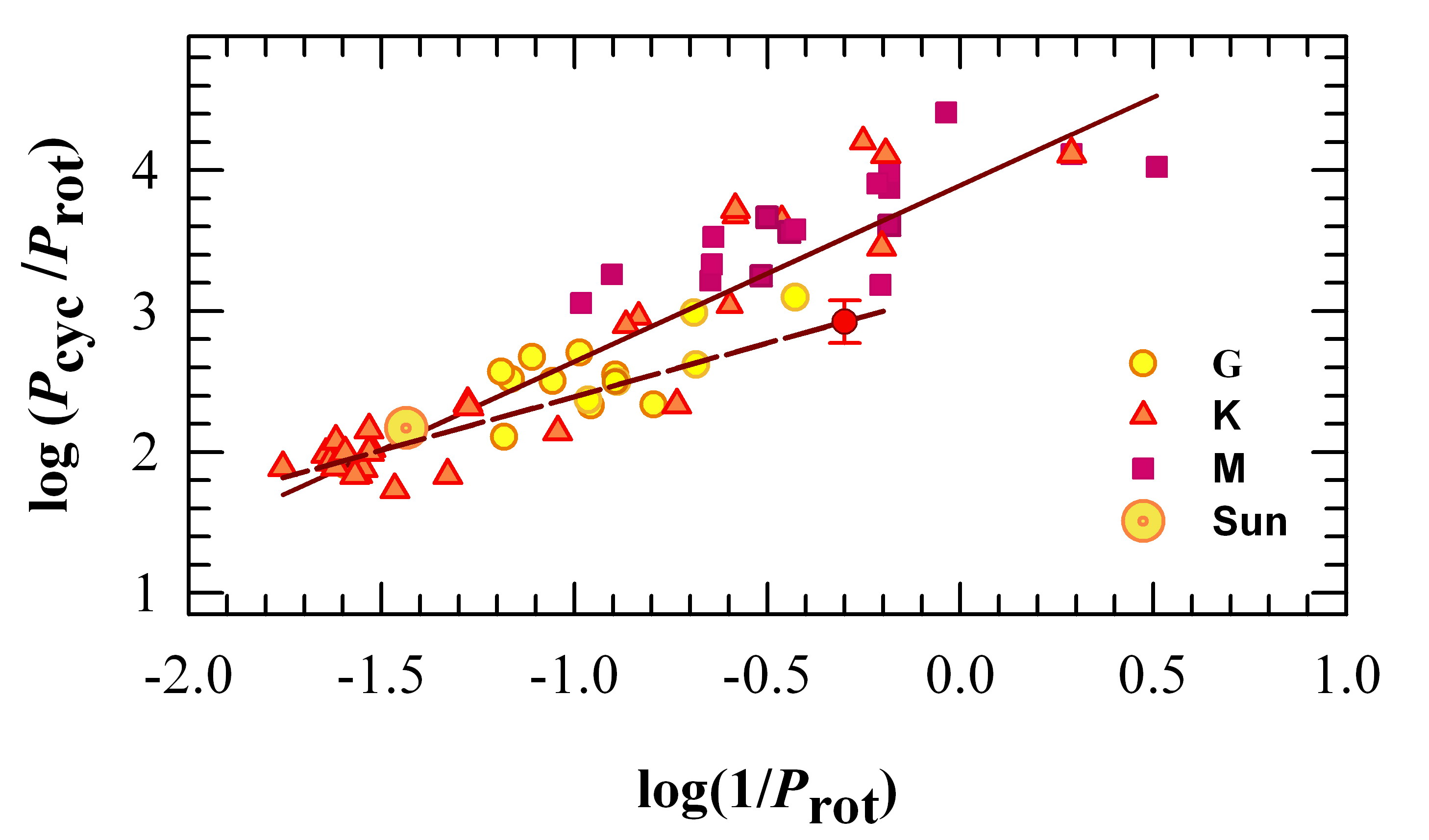}
\vspace*{0.2cm}
\caption{
Relation between the number of revolutions in the cycle and the rotation period for G--M dwarfs. The spectral classes of stars are indicated as in Fig.\ 2. The position of the Sun is shown by the corresponding symbol. The solid regression line fits all the stars of our sample, the dashed line shows the regression line from Oláh et al.~(2016).
}
\label{Figure4}
\end{figure*}

Figure 4 shows the essential difference between the slopes of regression lines for the entire sample of stars (solid line, $i = 1.25 \pm 0.068$) and for the dominant cycles of stars from Oláh et al.~(2016) (dashed line, $i = 0.76 \pm 0.15$). The differences in slopes are caused not only by possible uncertainties in the determination of cycle parameters, but are related to a large fraction of rapidly rotating dwarfs in our sample. However, there is still no a sufficient number of stars with $P_\textsl{rot} <5$~d to analyze such a relation for each spectral group.

\section{CONCLUSION}

The stars on or near the main sequence are related by structure (they have convective shells) and the observed manifestations of activity provided by the presence of magnetic fields. Different methods of observations on long-term time intervals reveal slow quasi-periodic changes in brightness of active dwarfs on intervals compared to the duration of the solar cycle. Such a pattern of changes is caused by the evolution of surface spots. For a number of stars, the analogy between the 11-year Schwabe cycle and the dominant cycles of stars is confirmed by the long-term observations. Multiple cycles, the butterfly effects were found in some stars studied in detail. The longer cycles, by several decades, are assumed to be caused by a change in spottedness of the stellar surface. Confident evidence for the evolution of spots on such time spans may be obtained from the long-term observations in different photometric bands. Modern methods for measuring magnetic fields make it possible to detect magnetic spots of different polarities, and from the many-year data one can identify cycles similar to the Hale cycle. The possibility of detecting such cycles on the stars YZ CMi, DT Vir, DX Cnc and others is discussed by Shulyak et al.~(2015). Differences in the duration of dominant cycles are associated with differences in the inner structure of stars, their physical parameters, and rotational periods. Certain results on the relation between characteristics of stars and their activity level were obtained for the solar-type stars (Oláh et al., 2016; Fabbian et al., 2017) from many-year observations of indicators of activity in the chromosphere and photosphere. Later stars are mainly investigated on time intervals of about 10 years.

On the basis of long-term observation series, features in the development of activity cycles were considered for 65 G--M dwarfs with rotation periods from fractions of a day to several tens of days. No cycles of more than 40 years were found in the solar-type stars; for most of them, activity cycles are in the range from 7 to 21 years, i.e. compared to the 11-year solar cycle, also having du-ration from 7 to 18 years at different epochs. Cycles of the fast-rotating dwarfs can reach 80 years. So far, such long cycles have been suspected in two stars -- V833Tau and V647 Her. About 30\%\ of the considered stars show only short dominant cycles of less than 7 years. 

For rapidly rotating dwarfs, there is no definite relationship between the cycle length and the rotational period, which is typical for stars with rotation periods longer than 5 days. Their cycle length increases with a deceleration of rotation. But in this group of stars, the longer cycles are observed in K-dwarfs rotating faster than the Sun.

\section*{ACKNOWLEDGEMENTS}

The authors are grateful to the referee for useful remarks and suggestions.

We used data from the following databases: ASAS and SuperWASP catalogs, the International Variable Star Index (VSX) database, the Kamogata – Kiso – Kyoto Wide-Field Survey data. The authors are thankful to all the staff for supporting these databases and providing access to data.

\section*{FUNDING}
This work was carried out with the partial support of the Russian Foundation for Basic Research (grants 18-52-06002 Az$_a$ and 19-02-00191a). 

\section*{CONFLICT OF INTEREST}
The authors declare that they have no conflicts of interest.

\phantomsection
\bibliographystyle{unsrt}

\end{document}